\documentclass[a4paper,11pt]{article}

\usepackage{amsmath}
\usepackage{amsfonts}
\usepackage{color}
\usepackage{xcolor}
\usepackage{url}
\usepackage{astats}
\usepackage{epsfig}

\def\bSig\mathbf{\Sigma}
\newcommand{\E}{{\mathbb{E}}}
\newcommand{\Jcal}{{\mathcal{J}}}
\newcommand{\Et}{{\widetilde{\mathbb{E}}}}

\renewcommand{\P}{\mathbb{P}}
\newcommand{\Pt}{{\widetilde{\mathbb{P}}}}

\newcommand{\N}{{\mathcal{N}}}

\newcommand{\Zt}{{\widetilde{Z}}}

\begin{document}
\title{Variational inference for coupled Hidden Markov Models applied to the joint detection of copy number variations}

\author{Xiaoqiang Wang$^{1,2*}$\url{xiaoqiang.wang@sdu.edu.cn}, 
Emilie Lebarbier$^{2}$, Julie Aubert$^{2}$, \\ and St\'ephane Robin$^{2}$ \\
$^{1}$School of Mathematics and Statistics, Shandong University, Weihai, China\\
$^{2}$UMR MIA-Paris, AgroParisTech, INRA, Universit\'e Paris-Saclay, 75005, Paris, France}

\maketitle

\begin{abstract}
Hidden Markov models provide a natural statistical
framework for the detection of the copy number variations (CNV) in
genomics. In this paper, we consider a Hidden Markov Model involving
several correlated hidden processes at the same time. When dealing
with a large number of series, maximum likelihood inference
(performed classically using the EM algorithm) becomes intractable.
We thus propose an approximate inference algorithm based on a
variational approach (VEM). A simulation study is performed to assess the performance of the proposed method and an application to the detection of structural variations in
plant genomes is presented.
\end{abstract}


\section{Introduction}
\label{s:intro}
\paragraph{CNV detection} 
Copy Number Variation (CNV) refers to the DNA sequence variation which increasing or decreasing the genomic segment of at least 50bp (\cite{ZMM15}).  As a typical form of structural variation, CNV generally consists of duplication, insertion or deletion events.  Since first studies in 2003-2004 (\cite{LHA03, IFR04, SLT04}), CNV have been prevalently discovered in the human genome (\cite{MZY14}). While most of the CNV analyses arise in human health,  some have been proved to  be associated with, or directly cause diseases or clinical phenotype variations (\cite{WSS13, ZMM15, CL16}). 
Due to their potential functional effect, thus possibly altering phenotypes/traits of interest, CNVs have also been intensively identified in many animal and plant species in the past few years. For example, some studies have also investigated the effect of CNVs on agronomical traits, such as the milk production traits (\cite{XCB14}), the growth traits in cattle (\cite{ZUX16}), the flowering time trait in maize (\cite{LRG15}). All of these CNV analysis in animal and plant species glean some preliminary insights into factors linked to the genomic selection.

\paragraph{Method for CNV detection} 
\cite{ACE11} review extensively the experimental platforms applied to CNV discovery and genotyping, which include two hybridization-based microarray technologies: array comparative genomic hybridization (CGH), Single Nucleotide Polymorphism (SNP) microarray, and sequencing-based  next-generation sequencing (NGS) technologies.  Depending on the data architecture resulted from above different platforms, a number of statistical methodologies and software tools have been developed  to detect CNV. Moreover, their performance to detect CNV are usually compared in the literature, for instance,  \cite{LJK05} for CGH, \cite{DSG10} and \cite{WYR09} for SNP array, \cite{PDS11} for cross-platform between CGH and SNP, \cite{ZWW13}, \cite{MTP12} and \cite{JC16} for NGS.  Among these methods, hidden Markov models-based (HMM) and segmentation-based algorithm are two main types of approaches.  Particularly, several studies were investigated to analyse simultaneously multiple individuals in CNV discovery. For instance, \cite{WCT08} and \cite{LLL16} propose some HMM-like methods which making use of family informations from parent-offspring trios. \cite{PLB11}, \cite{TKW10}, \cite{ZSJ10} and \cite{HZW16} propose some segmentation-like methods which focusing on detecting common or rare CNV regions across individuals.  These analyses applying on multiple individuals  attract our great attention, because they could obviously improve the accuracy of CNV estimation in comparison with analyzing typically single individual. We believe that the positive performance might rely on the fact that CNV is inheritable (\cite{SWT09}). Consequently, altered in the same loci across the individuals with common phylogenetic past, such as between offspring and either of parents in trios cas. These facts imply that the relatedness between individuals is a useful factor in CNV detection.

\paragraph{Measure of relatedness} Relatedness is a fundamental concept in genetic association studies, however there does not exist a common way to define them. \cite{AB09}  present that kinship is a central concept to measure pairwise genetic relatedness among individuals. The kinship coefficient $s_{ij}$ between two individuals $i$ and $j$ is the probability that an allele selected randomly from $i$ and an allele selected randomly from the same autosomal locus of $j$ are identical by descent (IBD). Nowadays, SNP marker-based relative kinship estimates have proven useful and accurate for quantitative inheritance studies in different populations. This genetic relatedness matrix is called also genetic similarity matrix by \cite{SB15}, in particular, the authors summarize  the SNP-based measure accounting for minor allele fraction (MAF) of the SNP by a series of formulates as:
$$
s_{ij}(\alpha) = \frac{1}{L}\sum_{t=1}^L \frac{(Z_{i,t}-2p_t)(Z_{j,t}-2p_t)}{\left[2p_t(1-p_t)\right]^\alpha},
$$
where $Z_{i,t}$ is the minor allele count (0, 1 or 2) of individual $i$, $p_t$ is the population MAF at the $t^{\text{th}}$ SNP and $\alpha$ takes some integer values. The performance for the case of \(\alpha=-1,0,1,2\) is compared in \cite{SB15}.
\paragraph{Coupled HMM and intractable likelihood} 
As an extension of HMM, coupled HMM (CHMM) model a system of multiple interacting processes, they take into consideration the interactions between variables in the latent space rather than observation space (\cite{RGR02}). Intuitively, CHMM has the ability to capture the relatedness between individuals in CNV discovery. In fact, CHMM have been applied in several fields such as speech recognition (\cite{NO03}), disease studies (\cite{SXT13}), health informatics (\cite{GSH16}), electroencephalogram analysis (\cite{ZG02}) and bioinformatics (\cite{CFN13}). 

CHMM is an incomplete data model for which the EM algorithm (\cite{DLR77}) is the most popular algorithm to maximize the likelihood. However, the exact inference in CHMM raises some computational issues. Indeed, when considering $Q$ status, $I$ individuals and $T$ observations for each individuals, CHMM is a HMM, the state space of which consists in all possible combinations of individual status. So the number of hidden states is $K := Q^I$ and the complexity of each E step of a regular EM algorithm is $K^2 T = Q^{2 I} T$, which becomes intractable when $K$ (that is, either $Q$ or $I$) becomes large. Many efforts have been made to manage this complexity, mostly by modeling the $K \times K$ transition matrix in a parsimonious way. (\cite{SJ99}) use a mixture form
$$
\P(S_{j,t}|S_{1:I, t-1}) = \sum_{i=1}^I \omega_{ij} \P(S_{j,t}|S_{i,t-1}),
$$
where $S_{j,t}$ represents the status of individual $j$ at position $t$ and $ \omega_{ij}$ can be viewed as mixing weights, or strength of effect of chain $i$ on chain $j$. This strategy reduces the number of transition parameters from one $Q^{2I}$ to  $I(I+1)Q/2$. 

\paragraph{Variational approximation for CHMM} 
In this paper, we try to keep the original form of CHMM even when $K$ is large and we use variational approximation to make the E step of the EM algorithm computationally tractable. The resulting variational EM (VEM) aims at maximizing a lower bound of the log-likelihood. VEM was first explicitly introduced in machine learning such as \cite{SJJ96},  now this approach has been routinely applied and generalized in many different ways. \cite{Jaa00} gives a brief introduction and \cite{WaJ08} provide a very complete overview. 

The variational approximation consists in seeking for some tractable distributions $\Pt(S)$ to replace $\P(S|X)$, the conditional distribution of hidden status $S$ given the observed data $X$, in E step. Such an approximation relies on two ingredients. We first need to choose a divergence measure between the true conditional distribution $\P(S|X)$ and the approximated one $\Pt(S)$ and the most popular is the K\"ullback-Leibler divergence. Then we need to define the class of distributions $\mathcal{Q}$ within which $\P(S|X)$ is searched for. Obviously, this second choice is critical as $\mathcal{Q}$ has to be as large as possible to make the approximation better but $\mathcal{Q}$ must also contain only distributions that are computationally manageable. In the context of factorial HMMs, \cite{GhJ97} introduced the distribution family of independent heterogeneous Markov chains to approximate the original distribution. Using this distribution class yields, during the E step, at preserving the within individual dependencies while neglecting the between individual dependencies. In this paper, we will adopt the same approximation type.

\paragraph{Joint CNV detection}
Although several methods considering multiple individuals have been derived in the literature, their relatedness/dependency relationship across individuals is fuzzy because of lack of informations on the status transition structure. Under the framework of CNV discovery, the genetic kinship matrix can be nowadays computed from the SNP genotyping data, therefore, we propose a novel CHMM accounting for this genetic relatedness between individuals. Our model is inspired by the fact that the closer the genetic relationship between any two individuals, the more likely their hidden status.

\paragraph{Paper outline}
In the following section we define the probabilistic model for CHMM which accounting for the genetic kinship matrix and in Section 3 we present algorithms for exact and variational inference and learning in CHMM. In Section 4 we describe simulation results comparing exact and approximate algorithms for learning on the basis of time complexity and model quality, next, confirming the necessity of taking kinship relationship into account in model. We also apply CHMM to a time series dataset consisting of 336 maize lines. We discuss several generalizations of the probabilistic model in Section 5.

\section{Model}
\label{s:mod}
We consider a set of $I$ individuals ($i = 1, \ldots, I$). For each individual, we observe a series of (microarray) measurements $X_i =
(X_{i,t})$, that is supposed to vary according to the status (copy number) of the individual at 'time' $t = 1,\ldots, T$ (position along the genome). We denote $(S_{i,t})_t$ the hidden process for individual $i$, where $S_{i,t}$ can take $Q$ different values  (e.g. $Q = 3$, $-1 =$ 'deletion', $0 =$ 'normal', $1 =$ 'amplification'). In this setting, the state space of the joint hidden process $(S_t)_t$, with $S_t =(S_{1,t}, \ldots S_{I,t})$, consists in $K := Q^I$ possible values.

\subsection{Emission distribution}
We assume that the observed data are all conditionally independent given the hidden process $S$, with respective conditional distribution:
$$
 (X_{i,t} | S_{i, t} = q) \sim \N(\mu_q, \sigma^2_q),
 $$
 where $\mu_q$ represents the mean value of state $q$. In the following, the means will be gathered in the vector $\mu = (\mu_q)_q$. We further denote $S_{i,t}^q = \mathbf{1}_{\{S_{i,t}=q\}}$ and $\phi_q(X_{i,t})$ the conditional probability density function of $X_{i,t}$ given the value $q$ of state $S_{i,t}$.

It is worth noting that dependent process was already considered in \cite{PLB11} in the same segmentation context. In this paper, the dependency was encoded in the joint distribution of the observed signals at each position $(X_{1,t}, \dots X_{i,t}, \dots, X_{I,t})$, making the normality assumption critical to achieve the inference. In our model, the dependency is encoded in the hidden layer so the emission distributions can be chosen arbitrarily. The Gaussian distribution is only chosen here to fit with microarray data. The same model could be easily adapted to sequencing data using Poisson or negative binomial distribution, or to any other type of signal.

\subsection{Hidden Markov chain}
\begin{figure}
\centerline{\includegraphics[width=.48\textwidth]{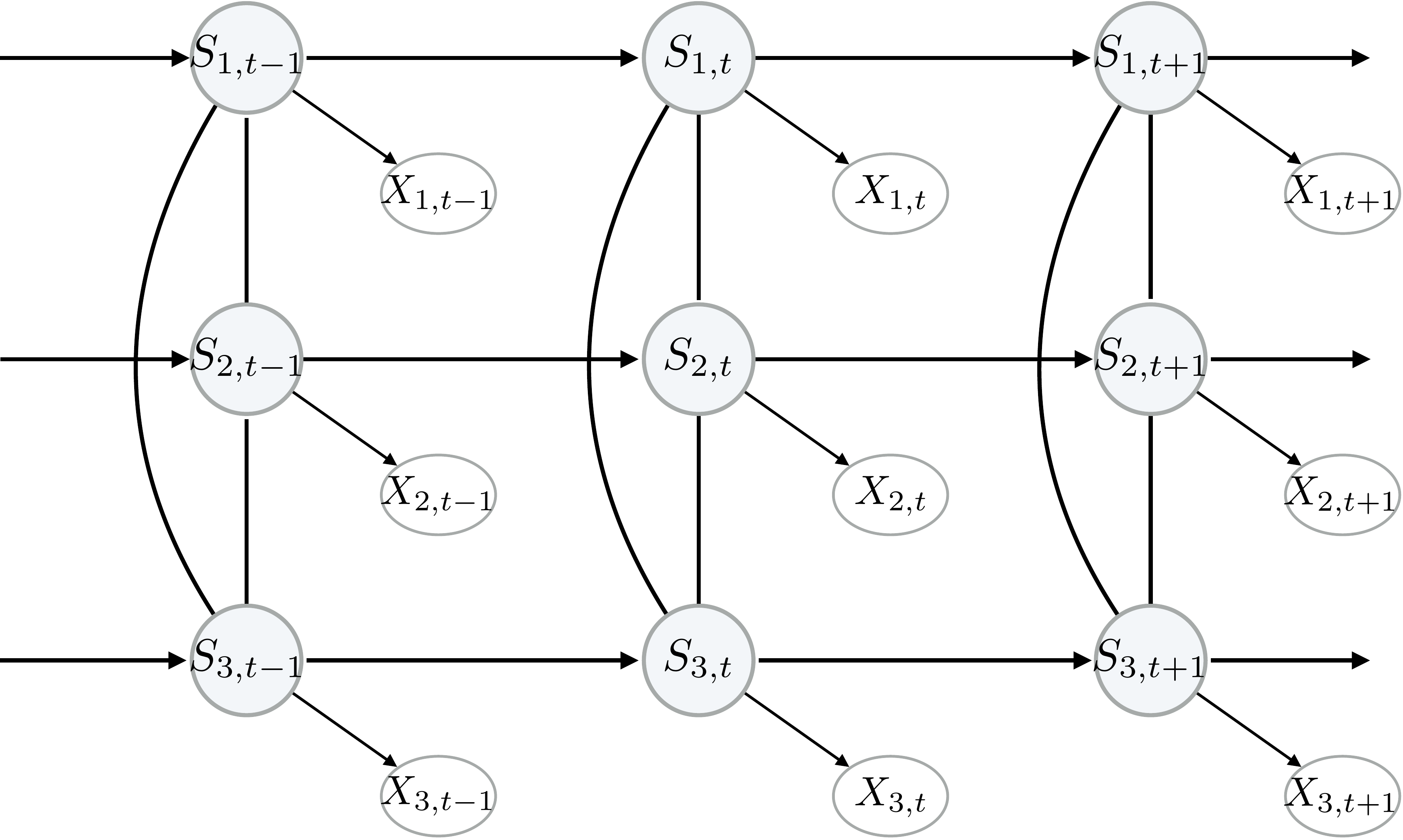}}
\caption{A mixture of directed and undirected graphical model. The directed edges represent the dependency relationship within the individual. The undirected edges represent the correlations among individuals. 
\label{fig:depstruc}}
\end{figure}

We also assume that joint hidden process $S$ is distributed according to a Markov chain. One purpose of our work is to introduce a hidden dependency structure as in Figure~\ref{fig:depstruc}. More specifically, the set of status of all individuals $(S_{i, t})_ i$ is a Markov chain and the edges between the status of all individuals at a given time $t$ allows to account for their (phylogenetic) proximity. Note that these edges introduce a coupling between the individual's hidden processes.

The variation of the hidden status along time as well as their correlation from one individual to another is encoded in the $K
\times K$ transition matrix $P$ of the joint hidden process $(S_t)$. We consider that the transition probabilities result from the
product of two terms: ($a$) one  accounting for the transitions within each individual and ($b$) one accounting for the similarities
between individuals (both supposed to be constant along time):
\begin{equation}
  \P(S_t = \ell | S_{t-1} = k) =: P_{k\ell} \propto \left(\prod_i \pi_{k_i, \ell_i} \right) W_\ell,
 \label{eq:Pkl}
\end{equation}
where
\begin{enumerate}
 \item[($a$)] $\pi$ is a $Q \times Q$ transition matrix (each row sums to one) and $k_i$ (resp. $\ell_i$) stands for the hidden state of individual $i$ when the joint hidden state is $k$ (resp. $\ell$); 
 \item[($b$)] the dependency relationship among the individuals is encoded in the coefficients
 \begin{equation} \label{eq:Wl}
  W_\ell = \prod_{i,j\neq i} \omega^{s_{ij}\mathbf{1}_{\{\ell_j\neq \ell_i\}}},  
 \end{equation}
 where $\omega  \le 1$ and $s_{ij}$ denotes the similarity (e.g. phylogenetic proximity) between individuals $i$ and $j$. Note that considering $\omega  = 1$ means considering the independent case.
\end{enumerate}
In this model, $\pi$ rules the within individual transitions, while $W$ introduces dependency between the individuals.

We further assume that the initial state $S_1 = (S_{i,1})_i$ has distribution
\begin{equation} \label{eq:initial_dist}
\P(S_1 = \ell) \propto \left(\prod_i m_{\ell_i} \right) W_\ell
 \end{equation}
where $(m_q)$ is a distribution of the states $1 \leq q \leq Q$.

Because the initial distribution \eqref{eq:initial_dist} and the transitions \eqref{eq:Pkl} are not normalized, the distribution of the hidden process $S$ writes
$$
\P(S) = \frac1Z \prod_{\ell} \left[\left(\prod_i m_{\ell_i} \right) W_\ell\right]^{S_1^\ell}
\prod_{\substack{t \geq 2\\k, \ell}} \left[ \left(\prod_i \pi_{k_i, \ell_i} \right) W_\ell \right]^{{S_{t-1}^k S_t^\ell}}
$$
where $Z$ stands for the normalizing constant. 
\footnote{It follows that
\begin{eqnarray*}
\log\P(X,S) &=& \sum_{i,q}S_{i,1}^q \log m_q + \sum_{i,t\geq 2,q,r}S_{i,t-1}^qS_{i,t}^r \log \pi_{q,r}\\
&&+\sum_{i,t,r}S_{i,t}^r\sum_{j\neq i}(1-S_{j,t}^r)s_{ij}\log\omega \\
&&+ \sum_{i,t,r}S_{i,t}^r\log\phi_r(X_{i,t}) - \log Z
\end{eqnarray*}
}

\section{Inference}
\label{s:inf}
 This section introduces the variational inference algorithm we propose. 

\subsection{EM algorithm}
The EM algorithm aims at estimating all the parameters, denoted $\theta$, for a fixed number of states $Q$ and a fixed parameter $\omega$. Indeed, $\omega$ could be estimated along with all other parameters, but this introduces some instability in the behavior of the algorithm. A heuristic to estimate $\omega$ outside the EM algorithm is given in Section \ref{sec:calibration_omega}. 

The EM algorithm alternates two steps:
\begin{itemize}
 \item\textbf{E-step:} evaluate the moments of the conditional distribution of the hidden variables $\P(S|X)$ for a current value of the parameter $\theta$, say $\theta^{h}$;
 \item\textbf{M-step:} update the parameter $\theta$ by maximizing the conditional expectation of the complete $\log$-likelihood with respect to  $\theta$ 
$$ 
\theta^{h+1}=\arg\max_{\theta} \E_{\theta^{h}}\left[\log \P(X, S; \theta) | X\right].
$$
\end{itemize}

In the case of HMM, the E-step can be achieved via a forward-backward recursion. This step is the critical point for the inference of the model described in Section~\ref{s:mod}. Indeed, we can face three situations:
\begin{enumerate}
 \item[1.] If we do not take into account  the phylogenetic dependency (i.e. if we set $\omega=1$ in \eqref{eq:Wl}), then the individuals' hidden processes are independent, so the E-step can be achieved using the standard forward-backward recursion for each individual.
 \item[2.] If we take into account the phylogenetic proximity but if both the number of individuals and the number of states are small, namely if $K = Q^I$ remains bellow few tens, the global model can be considered as one single HMM and the E-step can be achieved using the forward-backward recursion with complexity $\mathcal{O}(T K^2)$.
 \item[3.] If we take into account the phylogenetic proximity and if $K$ is too large, the complexity of the E-step becomes prohibitive, so some alternative has to be proposed.
\end{enumerate}
In the first two cases a regular EM can be used. Our work focuses on the third case.

\subsection{Variational EM algorithm}
We follow the line of \cite{Jaa00} and \cite{WaJ08} to derive our variational approximation. We first observe that, for any distribution $\Pt$, we have
\begin{align} \label{eq:lowerbound}
  \log \P(X) & \geq  \log \P(X) - KL\left[\Pt(S) || \P(S|X)\right] \nonumber \\
   & =  \Et \log \P(X, S) - \Et \log \Pt(S) =: \Jcal(X, \theta, \Pt),
\end{align}
where $\Et=\E_{\Pt}$ and $KL$ stands for the K\"ullback-Leibler divergence. The inference strategy then consists in maximizing the lower bound $\Jcal(X, \theta, \Pt)$ with respect to the parameter $\theta$. As EM algorithm, VEM alternates two steps:
\begin{itemize}
\item \textbf{VE-step:} update the approximate conditional distribution $\Pt$, given the current value of the parameter $\theta^{h}$, as
 $$
 \Pt^{h+1} = \arg\max_\Pt \Jcal(X, \theta^{h}, \Pt) = \arg\min_\Pt KL\left[\Pt(S) || \P(S|X; \theta^{h})\right].
 $$
 \item \textbf{M-step:} update the parameter estimates as 
 $$
 \theta^{h+1} = \arg\max_\theta \Jcal(X, \theta, \Pt^{h+1}).
 $$
 \end{itemize}

The quality of this approximation mostly relies on the class of approximating distributions within which $\Pt$ is searched for. We adopt here the general approach proposed by \cite{SaJ96} and adapted to the coupled HMM by \cite{GhJ97}, forcing $\Pt$ to be a product of independent Markov chains, that is
$$
\Pt(S) = \prod_i \Pt(S_i) 
\;\; \text{where}  \;\;
\Pt(S_i) = \prod_i \Pt(S_{i,1}) \prod_{t \geq 2} \Pt(S_{i, t} | S_{i, t-1}).
$$
We use the same parametrization setting
$$
\Pt(S_i) = \frac1{\Zt_i} \left(\prod_q (m_q h_{i1}^q)^{S_{i,1}^q}\right) \prod_{t \geq 2} \left(\prod_{q, r} (\pi_{q,r} h_{it}^r)^{S_{i, t-1}^q S_{i, t}^r}\right)
$$
where $\Zt_i$ stands for the normalizing constant ensuring that $\Pt(S_i)$ sums to one. The variational parameters $h_{it}^r$ can be viewed as correction terms with respect to a Markov chain with parameters $(m, \pi)$. 

We denote $\tau_{it}^r = \Et (S_{i,t}^r)$,  $\Delta_{it}^{qr} = \Et (S_{i,t-1}^q S_{i,t}^r)$ and 
$$
\log \Omega_{it}^r =  \left[\sum_{j \neq i} s_{ij} (1- \tau_{jt}^r)\right] \log \omega.
$$
Using the factorization properties of the approximating distribution $\Pt$, the lower bound $\Jcal(X, \theta, \Pt^h)$ given in \eqref{eq:lowerbound} becomes:
\begin{eqnarray*}
 \Jcal(X, \theta, \Pt^h) 
 & = & \sum_{i,r} \tau_{i1}^r [\log m_r + \log \phi_r(X_{i,1}) - \log (m_r h_{i1}^r)] \\
 & & + \sum_{i, t\geq 2, q, r} \Delta_{it}^{qr} [\log \pi_{q,r} - \log (\pi_{q,r} h_{it}^r)]\\
 & & + \sum_{i, t\geq 2, r} \tau_{it}^r [\log \Omega_{it}^r + \log \phi_r(X_{i,t})]\\
 & &  - \log Z + \sum_i \log \Zt_i \\
 & = & \sum_{i, t, r} \tau_{it}^r [\log \phi_r(X_{it}) + \log \Omega_{it}^r - \log h_{it}^r] \\
 & &- \log Z  + \sum_i \log \Zt_i,
\end{eqnarray*}
since $\Et(S_{it}^r S_{jt}^r) = \tau_{it}^r \tau_{jt}^r$ for all $i \neq j$ and since $\sum_q \Delta_{it}^{qr} = \tau_{it}^r$.

The VE-step consists in both finding the optimal value for the variational parameters $(h_{it}^r)$ and computing the approximate conditional moments $\tau_{it}^r$ and $\Delta_{it}^{qr}$. Following \cite{GhJ97}, Appendix D, we get
\begin{eqnarray*}
\frac{\partial \Jcal(X, \theta, \Pt^h)}{\partial \log h_{it}^r}  & = &    \Big[\log \phi_r(X_{i,t}) + \log \Omega_{it}^r - \log h_{it}^r \Big] \frac{\partial \tau_{it}^r}{\partial \log h_{it}^r}\\
& & -\tau_{it}^r + \tau_{it}^r,
\end{eqnarray*}
because $Z$ does not depend on $h_{it}^r$ and $\partial \log\Zt_i/ \partial \log h_{it}^r = \tau_{it}^r$. This derivative is zero for

\begin{equation} \label{eq:varparm}
h_{it}^r  = \Omega_{it}^r \ \phi_r(X_{i,t}).  
\end{equation}

The conditional moments, which depend on the normalizing constants $\Zt_i$, are then computed using an independent forward-backward recursion for each individual $i$:
\begin{itemize}
 \item Forward recursion: set $F_{i,1}^q \propto m_q h_{i1}^q $ and, for $t \geq 2$, compute
 $$
 F_{i,t}^r \propto \sum_q F_{i, t-1}^q \pi_{q,r} h_{it}^r;
 $$
 \item Backward recursion: $\tau_{iT}^r = F_{i,T}^r$ holds and, for $1 \leq t \leq T-1$, compute 
 $$
  G_{i, t+1}^r = \sum_q F_{i, t}^q \pi_{q,r} , \quad
 \Delta_{it}^{qr} = \pi_{q,r} \frac{\tau_{i, t+1}^{r}}{G_{i, t+1}^{r}} F_{i,t}^q, \quad
 \tau_{it}^q = \sum_r \Delta_{it}^{qr}.
 $$
\end{itemize}

\paragraph{Model selection} 
The number of states $Q$ can be fixed according to the considered problem as in our application study (Section \ref{s:app}). However, it can be difficult to choose in advance. We thus propose a criterion relying on the popular BIC criterion (\cite{Sch78}), which consists at subtracting from the maximized likelihood $\log \widehat{\P}(X)$  the penalty term $0.5 D \log(N)$ where $N$ is the number of observations and $D$ the number of free parameters. In our case, we have $N = I T$ and $D = 1 + Q(Q-1)$, so the penalty writes 
$$
\text{pen}_{BIC} = [1 + Q(Q-1)] \log(I T) / 2.
$$
Still, the likelihood of the observed data can not be computed in practice so we simply replace it by its variational lower bound and choose $Q$ as
$$
\widehat{Q} = \arg\max_Q \Jcal_Q(X, \widehat{\theta}, \Pt) - [1 + Q(Q-1)] \log(I T) / 2,
$$
where $\Jcal_Q(X, \widehat{\theta}, \Pt)$ is the maximized lower bound of the $Q$-state model (see e.g. \cite{DPR08}).

\subsection{Classification}

The aim of CNV analysis is to associate each genetic locus with a status, e.g. 'deleted', 'normal' or 'amplified'. So, the inference procedure requires a classification step that returns a predicted value for each $S_{i, t}$ to be completed. For a given locus $t$ in a given individual $i$, the VEM algorithm provides us with $\tau_{it}^r = \Pt\{S_{it} = r\}$ that is the variational approximate of $\P\{S_{it} = r | X\}$. A local classification rule then consists in simply taking the most probable status according to the $\tau_{it}^r$, that is to take
$$
\widetilde{S}_{it} = \arg\max_r \tau_{it}^r.
$$
Still, in many HMM applications, one is often interested in classifying all loci at once, which means retrieving the most probable hidden path $\widehat{S} = \arg \max_S \P(S | X)$. Because $\P(S | X)$ is intractable, we consider its variational approximation $\widetilde{S} = \arg \max_S \Pt(S)$, which can be obtained via the following Viterbi recursion. Let us denote \begin{align*}
  \alpha_{i,t}^r & = \max_{r_1,\ldots, r_{t-1}}\Pt(S_{i,1}=r_1, \ldots, S_{i,t-1} = r_{t-1}, S_{i,t}=r), \\
  p_{itqr} & = \Pt(S_{i,t}=r|S_{i,t-1}=q) \propto \pi_{q,r}h_{it}^r.
\end{align*}
At the first position of each profile, we have that $\alpha_{i,1}^r = \tau_{i1}^r$. Then, we apply the classical recursion 
$$
\alpha_{i,t}^r = \max_q \alpha_{i,t-1}^qp_{itqr}
$$
(for $t$ from 2 to $T$) and compute $\psi_{t}(r) = \arg\max_q \alpha_{i,t-1}^q p_{itqr}$. When reaching the last locus, we obtain $\widehat{S}_{i,T} = \arg\max_r \alpha_{i,T}^r$ and the rest of the optimal path is obtained recursively as $\widetilde{S}_{i,t} = \psi_{t+1}(\widetilde{S}_{i,t+1})$ (for $t$ from $T-1$ to 1).
\section{Simulation studies}
\label{s:sim}
To assess the performance of our approximated inference procedure, so-called here {\it CHMM-VEM} for Variational EM for Coupled HMM, we perform two simulation studies which aim is to show the advantage of our method in terms of both computational time and classification. In Study 1, we compare the computation time of {\it CHMM-VEM} to the exact version (the EM algorithm called here {\it CHMM-EM} for EM for Coupled HMM). In Study 2, we illustrate the importance of accounting for the dependency. To this aim, we consider an independent HMM, but in order to allow a fair comparison we assume moreover that the emission parameters are common among the series. In this case, the parameters can be estimated using an EM algorithm. We denote it {\it iHMM-EM} , which is equivalent to {\it CHMM-VEM} with $\omega=1$.

\subsection{Simulation design and quality criteria}
\paragraph{Simulation design}
In Study 1, we considered an increasing number of individuals $I \in \{2,3,4,5\}$, whereas we kept it fixed to $I=10$ in Study 2. For both studies, the length of the series was set to $T=1000$ and the number of hidden states was fixed to $Q=3$. 

We considered respectively the homoscedastic case and the heteroscedastic one for residual errors. In homoscedastic case, we used Gaussian emission distributions with respective means $-1$, $0$ and $1$, and we considered an increasing sequence of noise standard deviation: $\sigma \in \{0.3,1,1.2\}$. The difficulty of the detection problem increases with $\sigma$. In heteroscedastic case, we consider two configurations based on (a) a Maize dataset (\cite{BBB13}) (b) Illumina HumanHap550 array data (\cite{WLH07}). Chosen means and standard deviation values correspond to estimated HMM parameters. $(a)$ We used Gaussian emission distributions with respective means $-2$, $0$ and $2$
and associated $\sigma = 2, 0.25,2$; $(b)$ we used Gaussian emission distributions with respective means $-3.5$, $0$ and $0.68$ and associated $\sigma = 1.3, 0.2,0.2$.

The correlation term $W_\ell$ \eqref{eq:Wl} depends on both the similarities and the parameter $\omega$. Here, in order to mimic real data, we extracted the similarity matrix $(s_{ij})_{i,j}$ ($[I \times I]$) from the genetic kinship matrix of 336 maize lines given in \cite{BBB13}. For $\omega$, we consider two values corresponding to two levels of correlation between individuals: one case with moderate dependency (such that $\log{\omega} = -0.35$) and one case with weak dependency (such that $\log{\omega}=-0.2$). We simulated the hidden states in the following way:
\begin{enumerate}
\item[1.] we fixed central altered positions every 50 positions, i.e. at positions $25, 75, \ldots, 975$;
\item[2.] around each central altered position, we set a window with Poisson distributed length (with mean 15) so that alterations have various lengths;
\item[3.] for each window, we sampled the combination $\ell$ ($1 \leq \ell \leq K$) of individual status with probability proportional to $W_\ell$, so each alteration is not carried by every individual. 
\end{enumerate}

Each configuration $(\sigma,\omega)$ was simulated 100 times. For each simulated dataset, both {\it iHMM-EM} and {\it CHMM-VEM} were run and the loci were classified using the Viterbi algorithm. \\

\paragraph{Comparison criteria} To study the computational time, we measure it as the mean of  runtime in second. The forward-backward is written in C, the rest is implemented in R. We consider the classification between normal (0) and altered (-1 or +1) loci. To evaluate the performances, we use the following different criteria:
\begin{itemize}
\item False positive rate (FPR): the proportion of erroneously detected alterations among the normal status,
\item False negative rate (FNR): the proportion of erroneously estimated normal status among the alteration status, 
\item Accuracy: the proportion of correctly estimated status.
\end{itemize}
'Positive' corresponds to the two alteration status and 'negative' to the normal one. 

For each configuration, we consider the average of these criteria over the 100 simulations.

\subsection{Choice of the parameter $\omega$} \label{sec:calibration_omega}
\begin{figure}
\centering
\includegraphics[width=.48\textwidth]{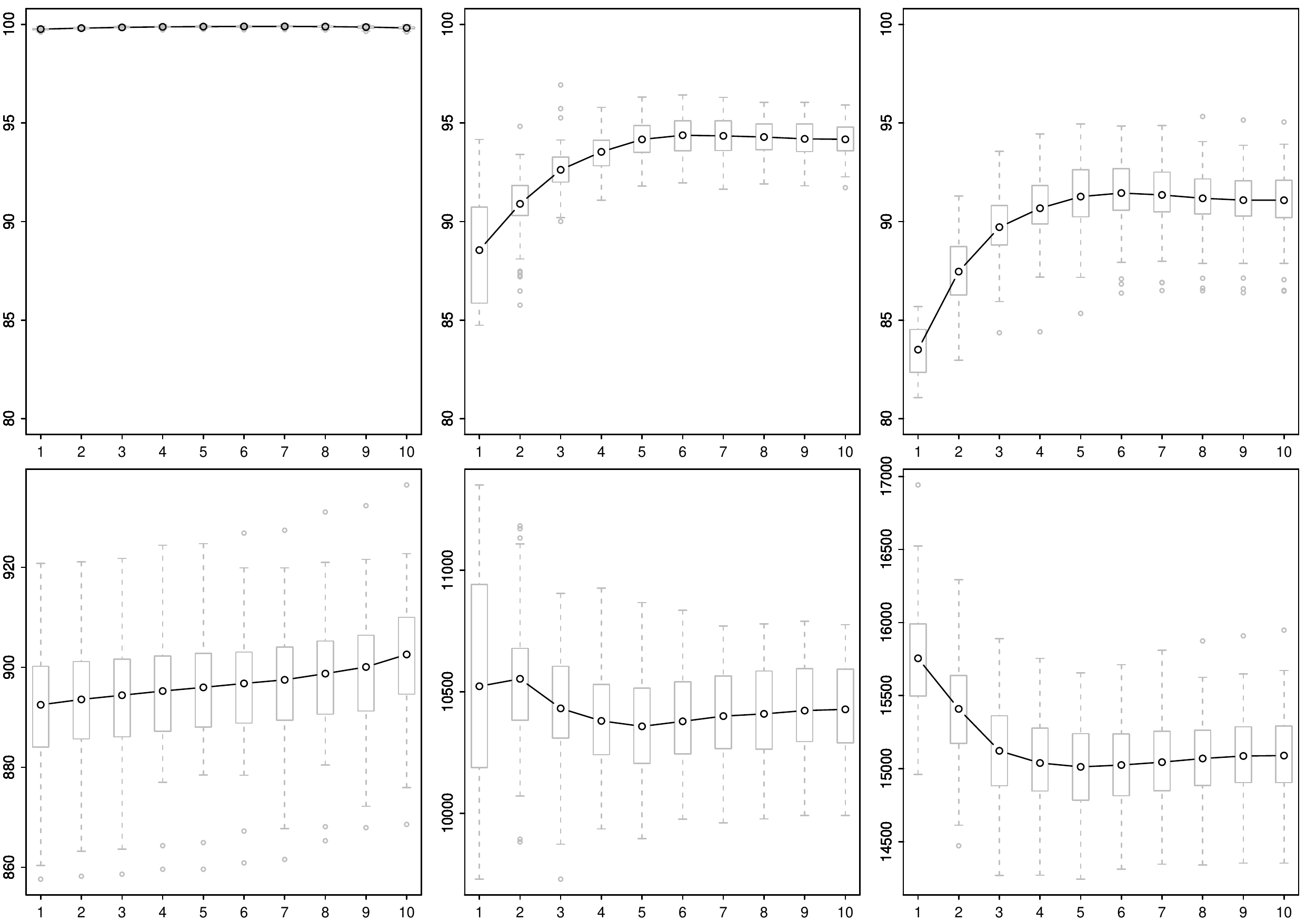}
\caption{Boxplot of accuracy (\%, top) and $\text{RSS}_{\omega}$ (bottom) for different values of $\omega \in \{e^{-k/20}| k=1,2,\ldots,10\}$. Left: $\sigma=0.3$, center: $\sigma=1$, right: $\sigma=1.2$.}
\label{fig:omgRSS}
\end{figure}

The proposed procedure does not allow to estimate the parameter $\omega$. To select it, we propose the following strategy: we vary $\omega$ in a grid of values and select the one that minimizes a weighted Residuals Sum of Squares ($\text{RSS}_{\omega}$) criterion defined by 
$$
\text{RSS}_{\omega} = \sum_{i,t,r} \tau_{it}^r(x_{i,t}-\mu_r)^2.
$$
Figure \ref{fig:omgRSS} gives both the classification accuracy and the $\text{RSS}_{\omega}$ criterion for different values of $\omega$ ($\log \omega = -k/20, k \in \{1, 2, \ldots, 10$\}) in the simulation case where $I=10$ and a weak dependency. Recall that a small value of $\omega \leq 1$ indicates a high dependency, so the x-axis of Figure \ref{fig:omgRSS} designs a decreasing level of dependency. We observe that when $\sigma$ is small, the accuracy is not affected by the choice of $\omega$ because the segmentation problem is obvious. For larger values of $\sigma$, we observe that the $\text{RSS}_{\omega}$ curve displays a minimum which is close to the maximum  classification accuracy. These phenomenons appear also in the case of the moderate dependency.

\subsection{Study 1} 
Only the results with weak dependency and  $\sigma = 1$ are presented, the other configurations lead to the same conclusions. Table~\ref{tab:time} gives the median of runtime in second on a PC with 3.2GHz with increasing number of individuals and Figure S1 
 in Supplementary presents the classification accuracy only with  $I=3$ individuals. As expected, {\it CHMM-EM} out-beats (slightly) {\it CHMM-VEM} followed by {\it iHMM-EM} in terms of accuracy. However, the runtime of {\it CHMM-EM} is exponential growth as $I$ increases and can not be used for larger (even small) number of individuals. Note that the runtime of {\it CHMM-VEM} is slightly better compared to {\it iHMM-EM}.
 
 \begin{table}[ht]
\centering
\begin{tabular*}{.48\textwidth}{*4{@{\extracolsep{\fill}}c}}
  \hline
$I$ & {\it iHMM-EM} & {\it CHMM-VEM} & {\it CHMM-EM} \\ 
  \hline
 2 & 0.8 & 0.4 & 2.0 \\ 
  3 & 1.1 & 0.5 & 11.2 \\ 
  4 & 1.2 & 0.6 & 79.4 \\ 
  5 & 1.6 & 0.8 & 920.2 \\ 
     \hline
\end{tabular*}
\caption{Runtime depending on the number of individuals $I$ (in second)}
\label{tab:time}
\end{table} 

Supplementary Figure S1 also shows that accounting for dependency between individuals improves the accuracy.

\subsection{Study 2}

\begin{figure}
\centering
\includegraphics[width=.48\textwidth]{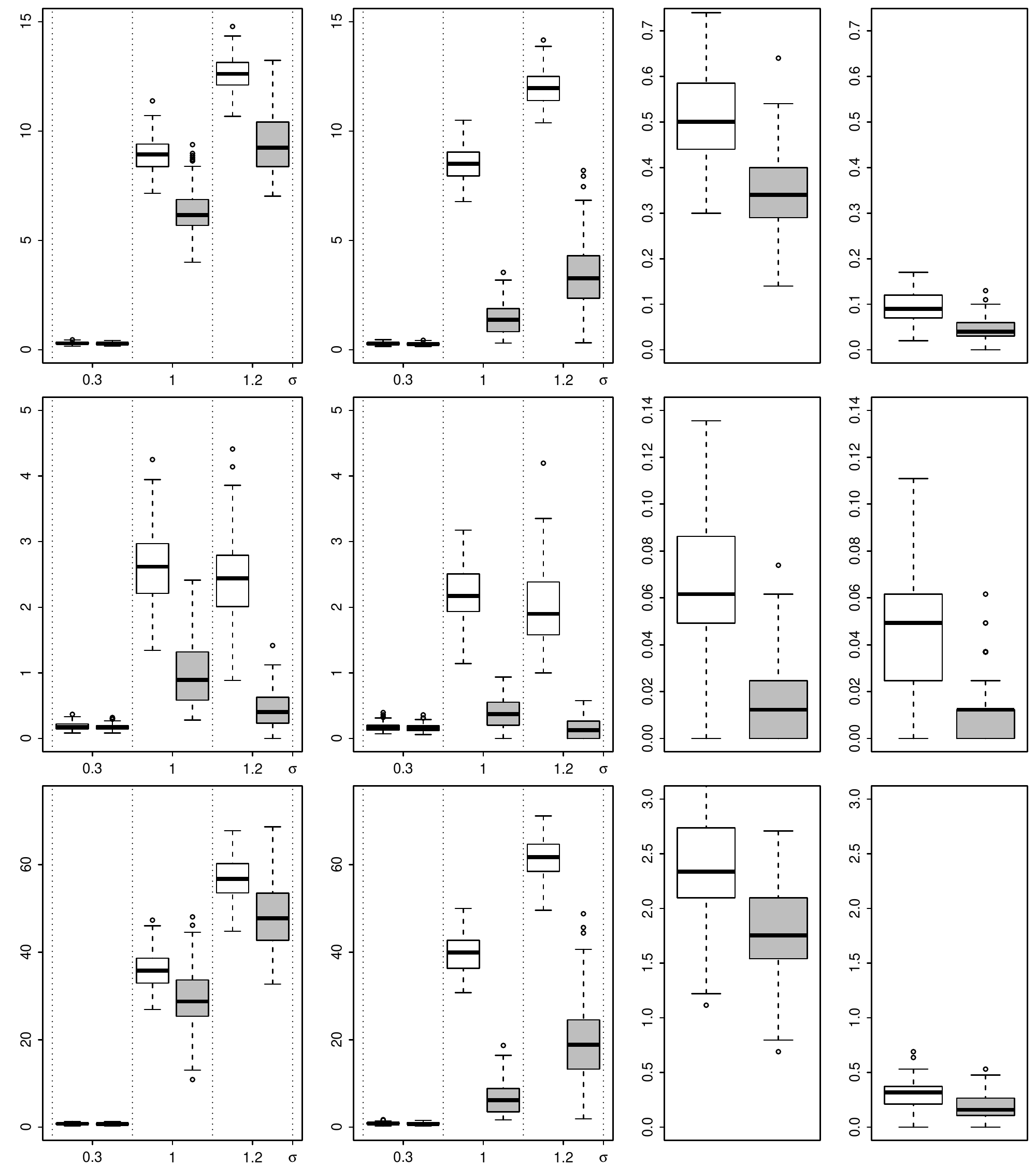}
\caption{First column: homoscedastic and weak dependency case. Second column: homoscedastic and moderate dependency case. Third column: first heteroscedastic case and weak dependency case. Forth column: second heteroscedastic case and weak dependency case. Boxplots of  classification error rate (\%, top), FPR (\%, center) and FNR (\%, bottom) for different values of $\sigma$ (x-axis). For each $\sigma$, we distinguish {\it iHMM-EM} (white box) and {\it CHMM-VEM} (gray box).}
\label{fig:VEMvsHMM}
\end{figure}

For each configuration, $\omega$ has been chosen following the strategy described in Section \ref{sec:calibration_omega}. In Figure~\ref{fig:VEMvsHMM}, we observe for homoscedastic model that when $\sigma$ small, i.e. when the detection problem is easy, both {\it iHMM-EM} and {\it CHMM-VEM} perform well. However, when $\sigma$ increases, {\it CHMM-VEM} outperforms {\it iHMM-EM} whatever the dependency, meaning the importance of taking into account for the existing dependency. This is more marked when this dependency increases. 

For heteroscedastic model, we observe from Figure~\ref{fig:VEMvsHMM} that {\it CHMM-VEM} outperforms {\it iHMM-EM} whatever the configurations.

\section{Application}
\label{s:app}
Maize is one of the three most cultivated crop in the world and a very interesting model for studying CNV and their impact on phenotype. CNV are very numerous in maize with thousand of CNV harboring hundred of functional gene between two inbred lines (\cite{LLX10, SEK10, SYF09}). \cite{LRG15} evaluated that one third of maize genome could be absent from B73 reference genome but present in another inbred lines.

Since the maize genotype B73 was sequenced in 2009 (\cite{SWF09}), B73 is usually considered as a reference genome to identify or understand the CNV in different types of maize. These experimental techniques have been investigated in different platforms such as in CGH platform (\cite{SEK10, BBH10, SYF09}); in NGS platform (\cite{WNY14, D16}).

\subsection{Data description}
We consider a dataset which consists of Illumina SNP genotyping arrays on $I = 336$ maize lines (\cite{BBB13}). The Illumina GenomeStudio software (see {\tt http://support.illumina.com/array/array\_software\\/genomestudio/downloads.html}) was used to compute the $\log$ R ratio (LRR) defined as 
$$
X_{it} = \log_2 \left( R_{it}^{\text{observed}} / R_t^{\text{expected}} \right)
$$
where $R_{it}^{\text{observed}}$ is the normalized signal intensity at locus $t$ in line $i$ and $R_t^{\text{expected}}$ is a reference intensity at locus $t$ (\cite{WLH07}). In this panel, two situations are expected: either the tested line $i$ shares locus $t$ with the reference genome and $X_{it}$ is close to zero (normal case) or locus $t$ does not exist in the genome of line $i$ and $X_{it}$ is below zero (altered case).

Among the 336 individuals, the French Fv2 inbred line has been especially studied and 58 deleted loci have been detected in contrast with B73 by sequencing method (\cite{D16}). 

\begin{figure}
\centering
\includegraphics[width=.48\textwidth]{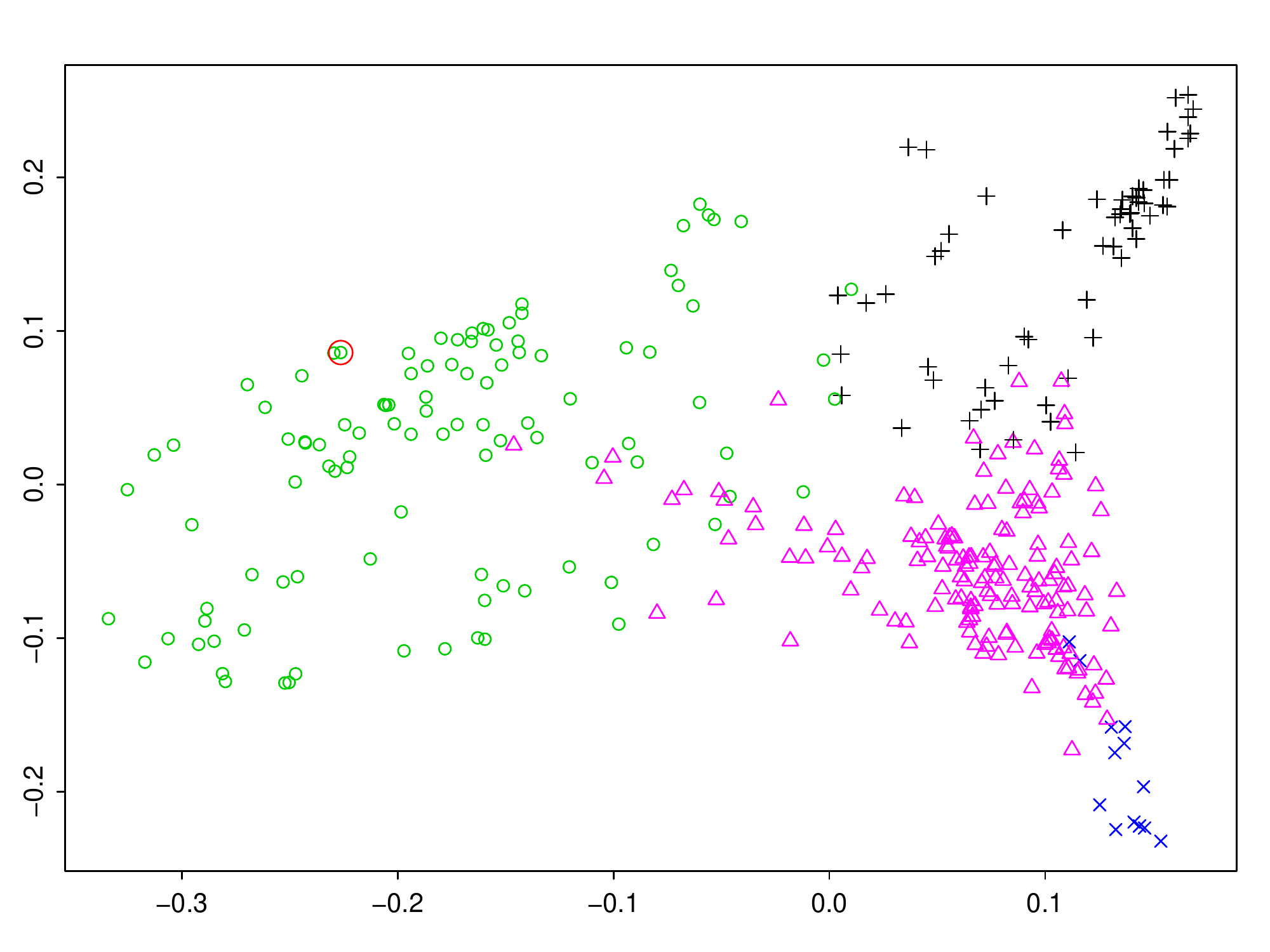}
\caption{Multidimensional scaling of the kinship matrix $(s_{ij})$: First two axes. Red circle: Fv2.}
\label{fig:ACPGr4}
\end{figure}

In addition to the Illumina array data, we have access to the kinship matrix $(s_{ij})$ between the lines (\cite{AB09}), which reveals the genetic similarity between them. Figure \ref{fig:ACPGr4} displays the multidimensional scaling (MDS) based on the similarity matrix. The clustering feature as shown in  Figure \ref{fig:ACPGr4} implies that we should analyze jointly the closed individuals rather than overall individuals.

\subsection{Data analysis}

\paragraph{Fixing the number of status}
As explained before, we expect only $Q=2$ status. To validate this, we fitted an HMM on each of the lines with $Q=2$, $3$ and $4$ status. Supplementary Figure S2  displays the histogram of the $Q\times I$ estimated means. Two main modes can be distinguished, which justify our choice of $Q=2$. 

Applying HMM to the analysis of 10 chromosomes in 336 individuals, we estimated the means as $-1.94(\pm 0.28)$ and $-0.05 (\pm 0.02)$ for two states, respectively. The corresponding variances of errors are estimated as $3.95(\pm 0.68)$ and $0.05(\pm 0.02)$(standard deviation $1.98(\pm 0.20)$ and $0.22(\pm 0.04)$).

\paragraph{Detecting CNV for 336 individuals}
As shown in Section~\ref{s:sim}, analyses jointly for correlated individuals are more effective than analyses independently from each other. Moreover, this effectiveness is obvious when the correlations among the individuals are strong.  In order to get some better correlated groups, we divide 336 individuals into 4 groups inspired from hierarchical clustering, then analyse one by one. The distance between these four groups is represented in Figure~\ref{fig:ACPGr4}. As shown in Supplementary Table S1, the analysis of the four groups compared to a single analysis gain nearly 1e4 deletion locus.

Supplementary Figure S3 displays the correlation between original similarity matrix and correlation matrix estimated separately by {\it iHMM-EM} and {\it CHMM-VEM}. We notice that the analysis accounting for the dependency between individuals by  {\it CHMM-VEM} are much more revealing than that of {\it iHMM-EM} in terms of similarity structure among individuals.

Supplementary Figure S4 lists the overlapped number of deleted loci for {\it iHMM-EM}  and {\it CHMM-VEM}. 

The simulation results from Figure~\ref{fig:VEMvsHMM} show that accuracy of {\it CHMM-VEM}  is greater than  that of {\it iHMM-EM} under some different parameter scenarios. Hence, we believe that {\it CHMM-VEM} gives more exact result than {\it iHMM-EM}, although {\it iHMM-EM} can find more deleted loci than {\it CHMM-VEM}.

\paragraph{Classification accuracy}
We use the 58 deletions detected in Fv2 by sequencing as references to compare the classification performances of {\it iHMM-EM} and {\it CHMM-VEM}. In particular, we study how the selection of the panel of lines does influence the results. To this aim, we ordered the lines by decreasing kinship with Fv2 and defined a sequence of panel with increasing sizes.

\begin{table}
\centering
\begin{tabular*}{.48\textwidth}{*7{@{\extracolsep{\fill}}l}}
  &  \it iHMM-EM & \multicolumn{5}{c}{\it CHMM-VEM}\\
  \cline{2-2}  \cline{3-7}
 					$I$  & 1        & 6        & 49 	& 80 		& 153	& 336 \\ \hline
   {$\overline s_I$}		     &  1.00  & 0.75  & 0.71 	& 0.67 	& 0.65 	& 0.64 \\ 
  				FPR(\%) & 12.68 & 10.43  & 10.02 	& 9.32   	& 8.89  	& 8.95\\ 
  				FNR(\%) & 24.14 & 24.14  & 24.14 	& 25.86 	& 25.86    & 25.86\\ 
   \hline
\end{tabular*}
 \caption{Classification accuracy of {\it iHMM-EM} and {\it CHMM-VEM}. $I:$ size of the panel. $\overline{s}_I:$ mean kinship within the panel. FPR and FNR for the validated 58 Fv2 alterations.}
\label{tab:Fv2Cor}
\end{table}

The results are given in Table \ref{tab:Fv2Cor}. We observe that the joint analysis with correlated lines reduces the proportion of falsely detected alterations.

\section{Discussion}
\label{s:discuss}
In practice, hundreds or thousands of individuals are often simultaneously analyzed to detect the CNV.  Especially for animal or plant species, these individuals share usually a common phylogenetic past. Therefore, their similarity relationship motivate us to focus firstly on constructing the probabilistic model on transition structure accounting for the kinship matrix. Next, we use the variational inference for CHMM in order to enable to handle jointly a large size of individuals. Simulation studies and real data analysis demonstrate that the account for the kinship between individuals improves the detection of CNV.

In addition, our transition models are compatible with the heterogeneous transition models and more sophisticated emission models such as in \cite{WLH07, SWT09, CYT07} in the context of CNV detection using SNP genotyping data. Furthermore, the read count data collected by NGS techniques is usually used to detect CNV in recent years.  Taking some emission distributions based on Poisson or negative binomial distribution such as \cite{WNY14, LLL16}, our model can be also easily extended to detect CNV for NGS platforms. 

Our method can be widened in CGH platform to detect CNV.That platform is based on the principle of comparative hybridization of two labelled individuals, say test and reference to a set of hybridization targets. The logarithm of signal ratio is used as the data to observe the copy number. For instance, when comparing two individuals test and reference, a deletion in  the reference individual is indistinguishable from an amplification in the test individual. 
In this section, as shown in Supplementary Figure S5, we consider a design which have multiple comparisons such as $(i,j), (i,k), (j,k), \ldots.$  We assume $m^+(i)$ the set of test individuals while taking individual $i$ as reference; conversely, we assume $m^-(i)$ the set of reference individuals while taking individual $i$ as test. Similar to the strategy in above sections, we search some independent heterogenous HMMs to approximate the original distribution as shown in  Supplementary Figure S5 in terms of K\"ullback-Leibler divergence $KL(\P||\Pt)$. Taking the same notations as above, the parameter in approximated HMM can be computed as 
$$
h_{it}^r = \Omega_{it}^r\prod_{j\in m^-(i), v} \phi_{\gamma_{rv}}(X_{(i,j),t})^{\frac{\tau_{jt}^v}{2}} \prod_{j\in m^+(i), u}\phi_{ \gamma_{ur}}(X_{(j,i),t})^{\frac{\tau_{jt}^u}{2}}
$$
(the product over all individuals compared with $i$ appears because the observations are paired, so we always have to deal with the joint distribution of $(X_{it}, X_{jt})$, as opposed as in \eqref{eq:varparm}).

The algorithms in current paper have been implemented in 
the R (\cite{R}) package {\tt CHMM}. The R package is available from the Comprehensive R Archive Network.


\section*{Acknowledgements}
This work was supported by the CNV-Maize program funded by the french National Research Agency
(ANR-10-GENM-104) and France Agrimer (11000415). Xiaoqiang Wang was financed by CNV-Maize project and National Natural Science Foundation of China (11601286).
We are grateful to St\'{e}phane Nicolas for providing the maize dataset.

\bibliographystyle{astats} 
\bibliography{Biblio.bib}

\appendix
\section*{Supplementary Materials}
\begin{figure}
\centering
\includegraphics[width=.48\textwidth]{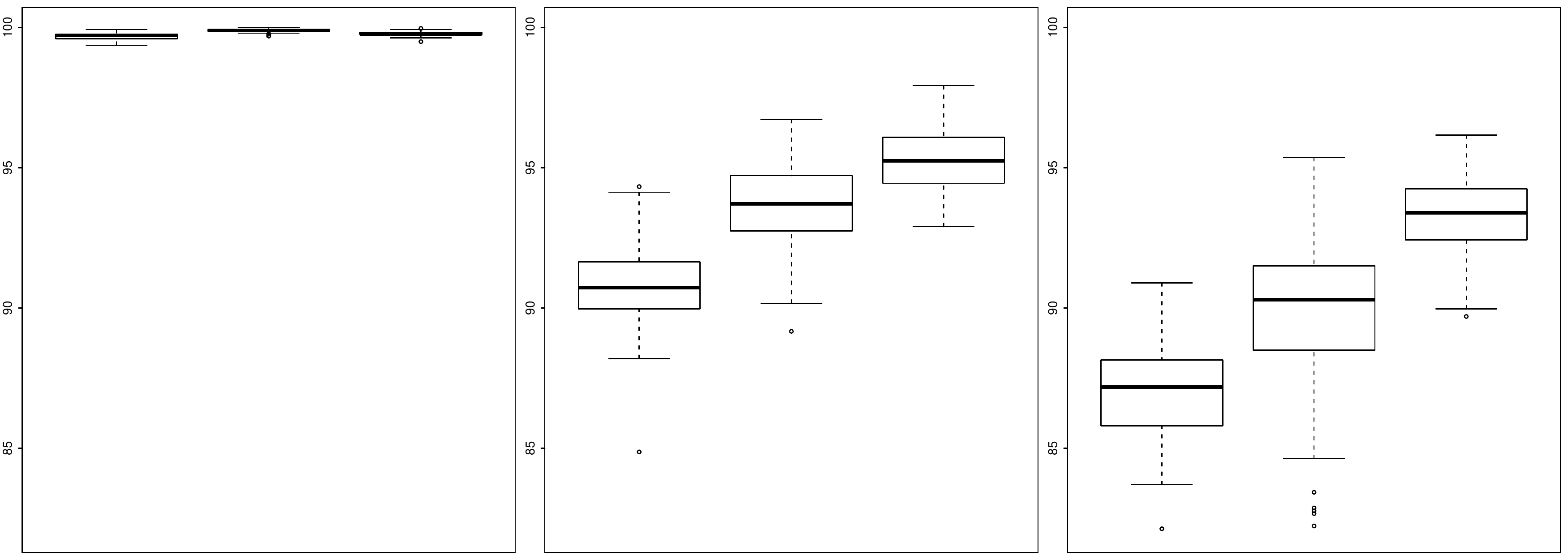}
\caption{Classification accuracy (\%) {\it iHMM-EM} (left), {\it CHMM-VEM} (middle) and {\it CHMM-EM} (right) for $I=3$. Left: $\sigma = 0.3$. Middle: $\sigma=1$. Right: $\sigma=1.2$}
\label{fig:acyComp}
\end{figure}

\begin{figure}
\centering
\includegraphics[width=.48\textwidth]{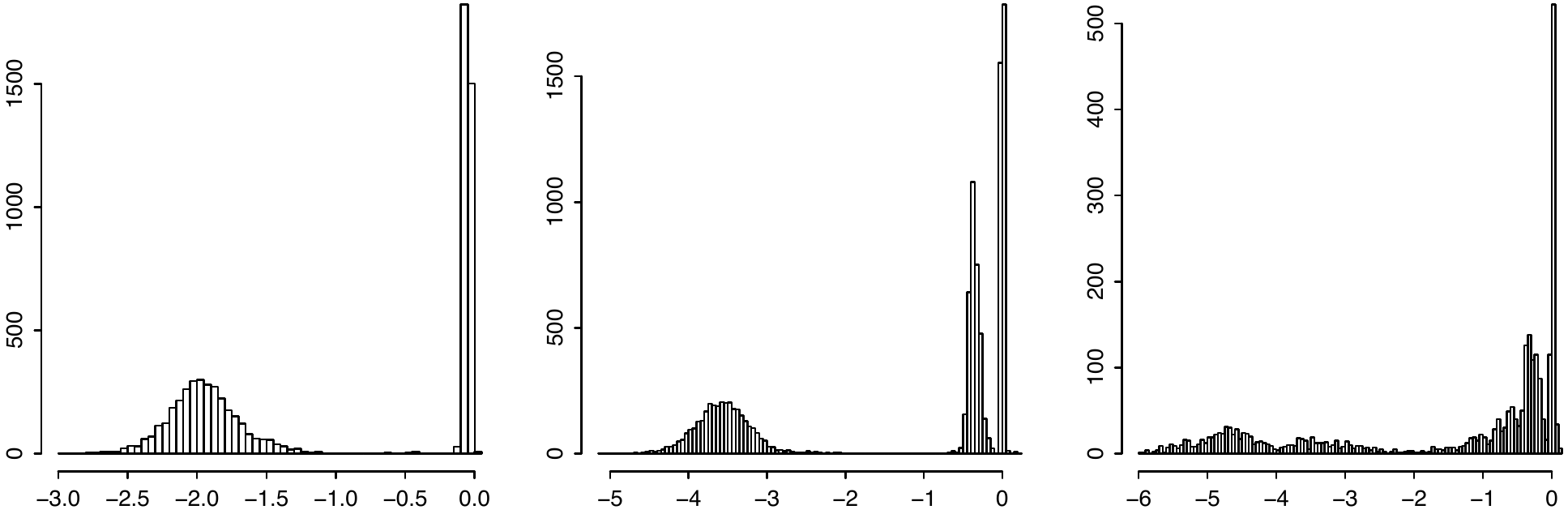}
\caption{Histogram of the mean estimated by independent HMM for 336 individuals. Left: $Q=2$, center: $Q=3$, right: $Q=4$, }
\label{fig:histMean}
\end{figure}

\begin{table}
\centering
\caption{Classification comparison between 4 groups and one 1 group}
\label{tab:group}
\begin{tabular*}{.48\textwidth}{*4{@{\extracolsep{\fill}}l}}
  \hline
  && \multicolumn{2}{c}{4 groups}\\ \cline{3-4}
  		&			& Deletion		& Normal\\\hline
1 group 	& Deletion		& 1469821 	& 49679 \\ 
		& Normal 		& 59456 		& 17082820 \\ 
   \hline
\end{tabular*}
\end{table}

\begin{figure}
\centering
\includegraphics[width=.48\textwidth]{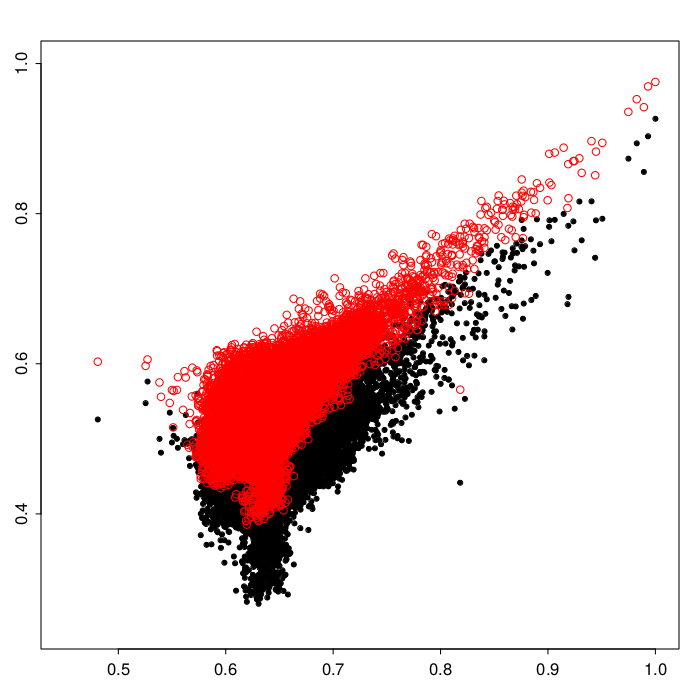}
\caption{Correlation between original similarity matrix and correlation matrix estimated by {\it iHMM-EM} (Black), {\it CHMM-VEM} (Red) }
\label{fig:corS}
\end{figure}

\begin{figure}
\centering
\includegraphics[width=.4\textwidth]{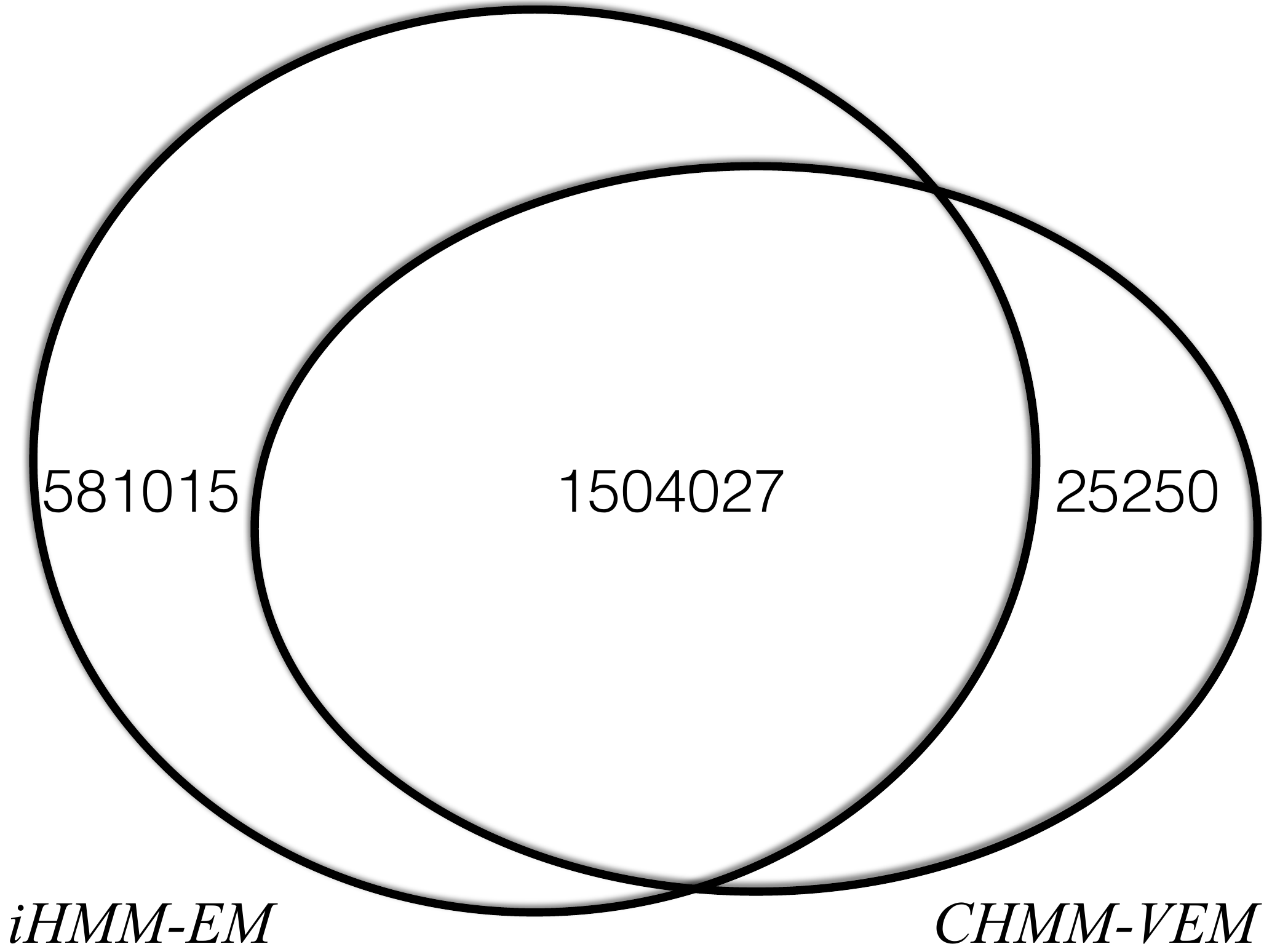}
\caption{Venn diagram of deleted loci detected by  {\it iHMM-EM} and {\it CHMM-VEM}}
\label{fig:overlap}
\end{figure}

\begin{figure}
\centering
\includegraphics[width=0.48\textwidth]{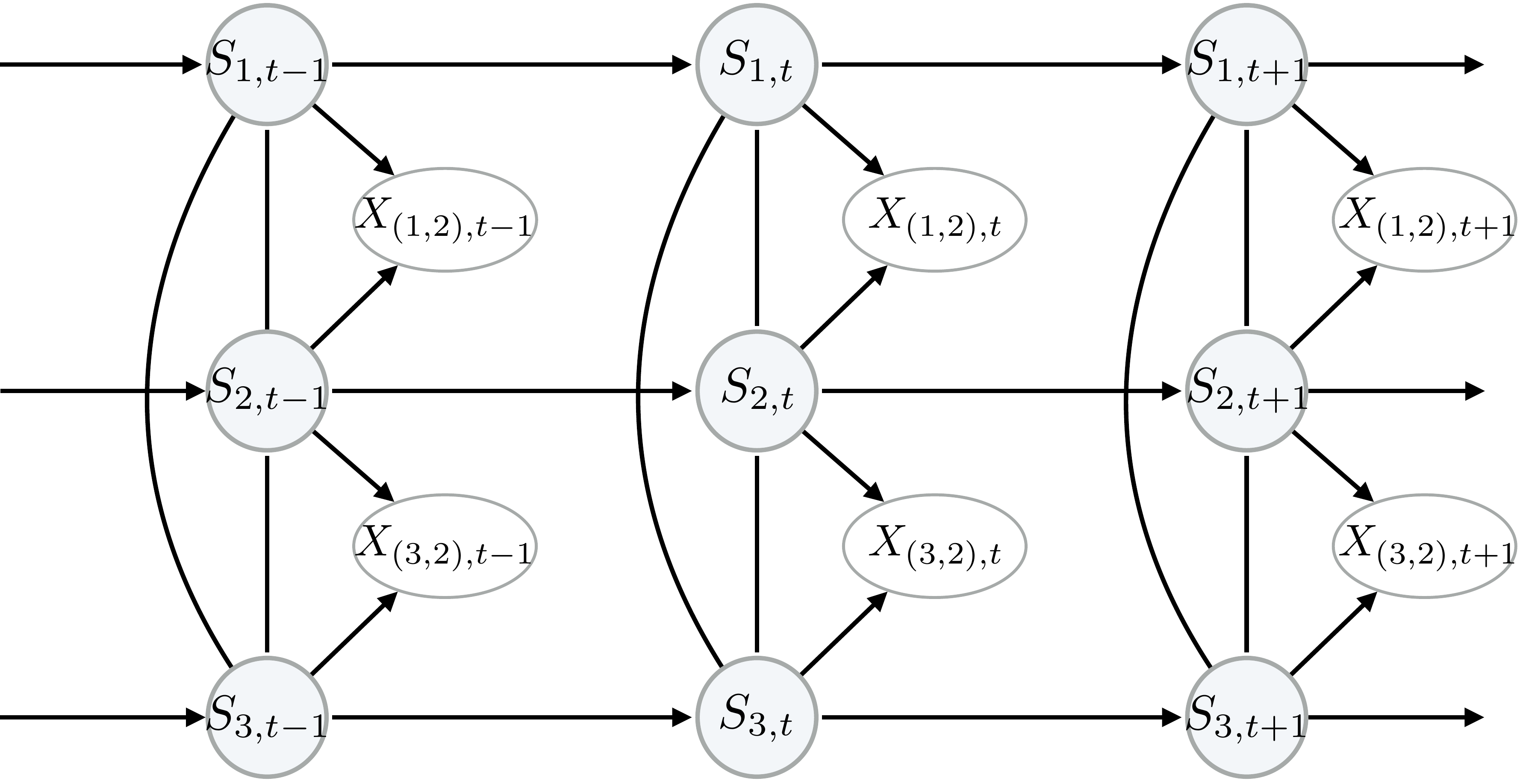}
\caption{A mixture of directed and undirected graphical model for CGH. The directed edges represent the dependency relationship within the individual. The undirected edges represent the genetical correlation among individuals.}
\label{Fig:CGH}
\end{figure}

\end{document}